\documentclass[12pt]{iopart}
\usepackage{graphicx}
\usepackage{xcolor}

\usepackage[LGR,T1]{fontenc}
\usepackage[latin9]{inputenc}
\usepackage{textcomp}
\usepackage{amssymb}
\usepackage{wasysym}
\usepackage{hyperref}
\usepackage{upgreek}
\usepackage{babel}

\usepackage{mathenv}

\usepackage{cite}
\usepackage{tikz}               

\begin{document}
\title[Bose-Einstein condensates and the thin-shell limit in anisotropic bubble traps]{Bose-Einstein condensates and the thin-shell limit in anisotropic bubble traps}

\author{Elias J. P. Biral$^1$, Nat\'{a}lia S. M\'{o}ller$^2$, Axel Pelster$^3$, F. Ednilson A. dos Santos$^4$ \\ \href{mailto:santos@ufscar.br}{santos.ufscar.br}}
\footnotesize{\hspace*{2.3cm} $^1$ Instituto de F\'{i}sica de S\~{a}o Carlos, Universidade de S\~{a}o Paulo, 13560-550 S\~{a}o Carlos,}\\
\footnotesize{\hspace*{2.3cm} SP, Brazil}\\
\footnotesize{\hspace*{2.3cm} $^2$ RCQI, Institute of Physics, Slovak Academy of Sciences, D\'{u}bravsk\'{a} Cesta 9,}\\
\footnotesize{\hspace*{2.3cm} 84511 Bratislava, Slovakia}\\
\footnotesize{\hspace*{2.3cm} $^3$ Physics Department and Research Center Optimas, Rheinland-Pf{\"a}lzische Technische}\\
\footnotesize{\hspace*{2.3cm} Universit{\"a}t Kaiserslautern-Landau, 67663 Kaiserslautern, Germany}

\footnotesize{\hspace*{1.6cm}$^4$ Department of Physics, Federal University of S\~{a}o Carlos, 13565-905 S\~{a}o Carlos,}\\
\footnotesize{\hspace*{2.3cm} SP, Brazil}

\begin{abstract}
\noindent Within the many different models, that appeared with the use of cold atoms to create BECs, the bubble trap shaped potential has been of great interest. 
However, 
the relationship between the physical parameters and the resulting manifold geometry remains yet to be fully understood for the anisotropic bubble trap physics in the thin-shell limit. 
In this paper, we work towards this goal by showing how the parameters of the system must be manipulated in order to allow for a non-collapsing thin-shell limit. In such a limit, a dimensional compactification takes place, thus leading to an effective 2D Hamiltonian which relates to up-to-date bubble trap experiments. At last, the resulting Hamiltonian is perturbatively solved for both the ground-state wave function and the excitation frequencies in the leading order of deviations from a spherical bubble trap.
\end{abstract}
\section{Introduction}
In the 1990's, the experimental realization of Bose-Einstein Condensates (BEC) \cite{BEC1, BEC2} gave rise to a myriad of both theoretical and experimental studies with contributions ranging from a basic understanding of the underlying physics of this macroscopic quantum phenomenon to various applications in particular cases of interest. Among the vast knowledge developed, there is the creation of bubble trap physics \cite{bubble0, bubble1, bubble2,Lundblad1,Salasnich1}, which consists of thin-shell traps created using a radiofrequency field in an adiabatic potential based on a quadrupolar magnetic trap.

The idea to work with two-dimensional superfluid manifolds soon proved to be appealing to physicists since the fine tune in the geometry opens new possibilities of physical interest. As a natural consequence, many experiments appeared in the literature \cite{bubble3, bubble4, bubble5, bubble6, bubblenovo}. Unfortunately, there are various technical difficulties in creating bubble trap experiments among which is the gravitational sag, i.e., the sinking of the BEC atoms into the bottom of the trap. With the current developments, it is possible to escape this problem working with microgravity either with free-falling experiments on earth-based laboratories \cite{bubble7, bubble8} or space-based in the International Space Station (ISS) with the Cold Atom Laboratory (CAL) \cite{CAL1, CAL2, CAL3, CAL4, CAL5, CAL6, CAL7}. Up until today, the usual microgravity \cite{microgravity} seems to be the best solution for confining atomic gases into shells in order to study its properties, but some new alternatives such as gravity compensation mechanisms are arising \cite{antigravity1, antigravity2}. Also, an interesting substitute to the usual procedure of radio-frequency dressing was proposed for dual-species atomic mixtures \cite{axel1}, which led to the creation of a BEC on Earth's gravity \cite{axel2}.

Confinements in three-dimensional shell shaped condensates inspired some theoretical works worth mentioning here.  For instance, in \cite{bubbleteo9}, the authors apply analytical methods to investigate the ground state wave function of a BEC and its collective modes. In \cite{bubbleteo1}, both analytical and numerical techniques are used in order to obtain expansion properties. The interesting paper \cite{bubbleteo2} employs thermodynamic arguments in order to survey the formation of clusters. The thermodynamics of a BEC on a spherical shell is analyzed, including the critical temperature, in Refs.~\cite{bubbleteo3, bubbleteo4, bubbleteo5}. The topological hollowing transition from a full sphere to a thin-shell was studied in \cite{bubbleteo6} and \cite{bubbleteo7}, thus finding some universal properties. The ground state and collective excitations of a dipolar BEC was considered in \cite{bubbleteo8}.  The general physical relevance of cold atoms on curved manifolds is also addressed in \cite{bubbleteo10}. The contribution of \cite{bubbleteo11} plays with the idea that the external potential is equivalent to the harmonic trap for a large radius thin-shell. Universal scaling relations are found for topological superfluid transitions in bubble traps in \cite{bubbleteo12}. And non-Hermitian phase transitions are meticulously described in \cite{bubbleteo13}. Although it is not the focus of this work, it is also worth citing vortex studies on spherical-like surfaces with no holes \cite{vortex1, vortex2, vortex3, vortex4, vortex5, vortex6, vortex7, vortex8}.

In this paper, we work out an explicit relation between the confinement of the particles in the thin-shell limit and the geometrical distortion of a bubble trap for a family of confinement potentials.
They are chosen in such a way that the current experiments are included as special cases. All potentials in this family turn out to exhibit the same angular dependency for the confinement strength.
Section 2 describes the mathematical background by defining the concepts in which our theory is developed.
In a second step we calculate in section 3 the harmonic radial frequency in the Gaussian Normal Coordinate System (GNCS) and expose its resulting angular dependency. Furthermore, the definition of the thin-shell limit is discussed in a more rigorous way by elucidating how it depends on the geometrical distortion of the bubble trap.
The general Gross-Pitaevskii Hamiltonian of the system is deduced using a perturbative approach near the thin-shell limit in section 4. In section 5, special topics of the spherical shell and the Thomas-Fermi approximation are considered. Finally, in section 6, the corresponding effective Gross-Pitaevskii equation is perturbatively solved
for small distortions from a spherical bubble trap in order to determine both the ground-state wave function and the oscillation
frequencies.
\section{Gaussian Normal Coordinate System }
In this section, some preliminary concepts are defined and explained in order to establish the mathematical background in which our theory is developed.
One of such main concepts concerns the manifold \cite{manifold1, manifold2} considered here. In this work, we study 2D surfaces embedded into a 3D Euclidean space. More specifically, ellipsoidal surfaces \cite{ellispoidal1, ellispoidal2, ellispoidal3} are considered since they correspond to the bubble trap potentials in BEC experiments.  Therefore, our manifolds are compact, smooth and differentiable everywhere. In order to describe the 3D region around the 2D manifold we choose as a suitable coordinate system, the so-called Gaussian Normal Coordinate System (GNCS) \cite{GNCS1,GNCS2}. Further features and particularities on its application can be found in \cite{bubbleteo9}.

It is always possible to describe the region around smooth manifolds with the aid of a GNCS. The main idea is to consider two coordinates $x^{1}$ and $x^{2}$ over the 2D manifold $M$, also called tangent coordinates, describing arbitrary points in the manifold. Thus, any point $\mathbf{p}$  of this manifold $M$ is portrayed by the position vector $\overrightarrow{p}(x^{1},x^{2})$. Any point $\mathbf{q}$ in such a vicinity of $M$ can be represented by a coordinate $x^{0}$ referred to as the orthogonal coordinate, and a normal unit vector $\hat{n}$ at the point $\mathbf{p}$ through the following equation
\begin{eqnarray}
\overrightarrow{q}(x^{0},x^{1},x^{2})=\overrightarrow{p}(x^{1},x^{2})+x^{0}\hat{n}(x^{1},x^{2}).\label{eq1}
\end{eqnarray}
We define the geometrical shape of the manifold in question with the prolate spheroidal coordinates \cite{prolate1, prolate2}. The transformation equations between such a system of coordinates and the Cartesian System allows us to establish the following family of ellipsoidal surfaces
\begin{eqnarray}
  \left\{\begin{array}{r@{}l@{\qquad}l}
	x= & A\sin\nu\cos\phi\\
	y= & A\sin\nu\sin\phi\\
	z= & \frac{A}{\sqrt{1+\epsilon}}\cos\nu
  \end{array}\right.
,\left\{\begin{array}{r@{}l@{\qquad}l} 
\phi\in[0,2\pi)\\
\nu\in[0,\pi) 
 \end{array}\right. .
\end{eqnarray}
Here $\nu = x^1$ and $\phi = x^2$ represent the tangent coordinates, whereas the parameter $\epsilon$ stands for the geometrical
distortion between an ellipsoid and a sphere according to the equation $x^{2}+y^{2}+(1+\epsilon)z^{2}=A^{2}$, where $A$ denotes a quantity analogous to a sphere radius characterizing the overall size of the ellipsoid.

For each point on the manifold $M$ it is possible to determine a pair of mutually orthogonal vectors tangent to the manifold by taking partial derivatives of $\overrightarrow{p}(\nu,\phi)$
\begin{eqnarray}
  \left\{\begin{array}{r@{}l@{\qquad}l}
	\vec{v}_{1}(\nu,\phi)= & \frac{\partial\overrightarrow{p}(\nu,\phi)}{\partial\nu}=A\cos\nu\cos\phi\hat{x}+A\cos\nu\sin\phi\hat{y}-\frac{A}{\sqrt{1+\epsilon}}\sin\nu\hat{z},\\
	\vec{v}_{2}(\nu,\phi)= & \frac{\partial\overrightarrow{p}(\nu,\phi)}{\partial\phi}=-A\sin\nu\sin\phi\hat{x}+A\sin\nu\cos\phi\hat{y}.\\
  \end{array}\right.
\end{eqnarray}
In this way the unitary normal vector can be written as
\begin{eqnarray}
\hat{n}=\frac{\vec{v}_{1}\times\vec{v}_{2}}{\left|\vec{v}_{1}\times\vec{v}_{2}\right|}=\frac{\sin\nu\cos\phi\hat{x}+\sin\nu\sin\phi\hat{y}+\sqrt{1+\epsilon}\cos\nu\hat{z}}{\sqrt{1+\epsilon\cos^{2}\nu}}.
\end{eqnarray}
These vectors form an orthogonal basis according to the following properties
\begin{eqnarray}
  \hat{n}\cdot\hat{n}= 1,\quad \vec{v}_{1}\cdot\hat{n}=  0, \quad	\vec{v}_{2}\cdot\hat{n}= 0, \quad	\vec{v}_{1}\cdot\vec{v}_{2}=  0.\label{properties}
\end{eqnarray}
Choosing $x^{0}=s$, the set of coordinates $(x^0,x^1,x^2)=(s,\nu,\phi)$  defines the GNCS.

\begin{figure}[t]
\centering
\graphicspath{{./Figuras/}}
\includegraphics[width=0.4\linewidth]{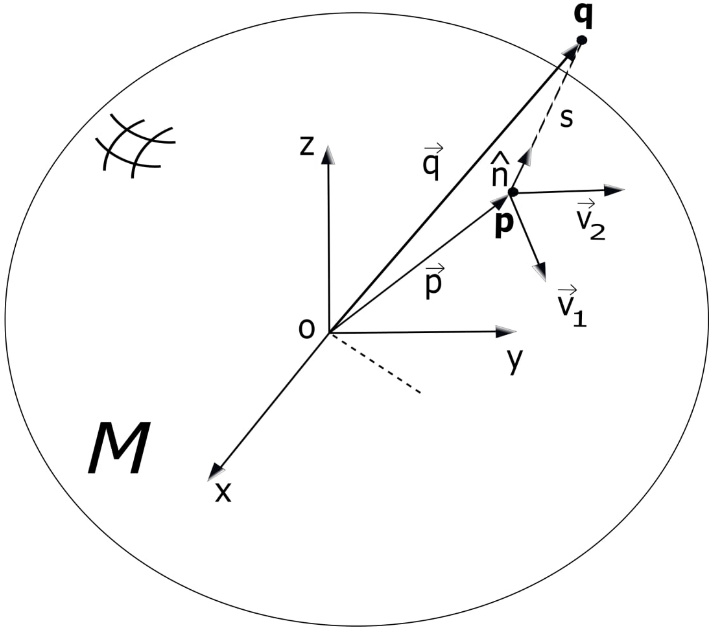}
\caption{Drawing of the manifold $M$ as an ellipsoid including the GNCS with  $(x^0,x^1,x^2)=(s,\nu,\phi)$ described with the aid of the prolate spheroidal coordinates developed in this section. }
\label{figure1}
\end{figure}

The visualization of the coordinates and vectors outlined in this section is exposed in Figure \ref{figure1}. It shows a drawing
of the manifold $M$ as an ellipsoid including the GNCS. 
The vector $\overrightarrow{p}(\nu,\phi)$ points to the point $\mathbf{p}$
at the manifold, where $s=0$. Also the two tangent vectors $\vec{v}_{1}(\nu,\phi)$
and $\vec{v}_{2}(\nu,\phi)$, as well as the unit normal vector $\hat{n}(\nu,\phi)$
to the manifold at the point $\mathbf{p}$ are illustrated.

Now, let us consider the 3D metric tensor $G_{\alpha\beta}=\frac{\partial\overrightarrow{q}(s,\nu,\phi)}{\partial x^{\alpha}}\cdot \frac{\partial\overrightarrow{q}(s,\nu,\phi)}{\partial x^{\beta}}$, which has in matrix notation according to the properties (\ref{properties}) the typical form within GNCS
\begin{eqnarray}
G_{\alpha\beta}=
\left[
\begin{array}{ccc}
1 & 0 & 0 \\
0 & \Bigg(\cos^{2}\nu+\frac{\sin^{2}\nu}{1+\epsilon}\Bigg)\Bigg(A+s\frac{1+\epsilon}{(1+\epsilon\cos^{2}\nu)^{3/2}}\Bigg)^{2} & 0 \\
0 & 0 & \sin^{2}\nu\Bigg(A+s\frac{1}{\sqrt{1+\epsilon\cos^{2}\nu}}\Bigg)^{2} 
\end{array} 
\right].
\label{metric}
\end{eqnarray}
It is important to realize that in the case of a spherical shell, with $\epsilon=0$, we recover the result of the metric tensor
for spherical coordinates, where $r = A+s$ denotes the radial coordinate and $\nu$ stands for the polar angle.
\section{Thin-shell limit for bubble traps}
In this section, we discuss the types of potentials that are relevant to this work. Namely, we consider a family of 3D potentials which are constant and have their lowest value along the manifold $M$, and that have their confinement strength proportional to the geometrical distortion of the ellipsoid. Later we show how such conditions are consistent with actual experimental potentials. 
\subsection{Confinement potential in a bubble trap}
The initial idea is to introduce a parameter $\Lambda$, which controls the strength of the confinement in the direction perpendicular to the manifold. To this end, we simplify the notation according to $(x^{0},x^{1},x^{2})\equiv (s,x^{i})$ with $i=1,2$, so the general potential can be considered as
\begin{eqnarray}
V(s,x^{i}) & =\Lambda \textrm{\texttwosuperior}v(s,x^{i}),\label{genpot}
\end{eqnarray}
where the limit $\Lambda\rightarrow\infty$
corresponds to an infinitely tight potential thus defining the thin-shell limit. In addition, the factor $v(s,x^{i})$ does not depend on the shell thickness and must be chosen in such a way that it fulfills the requisites of being constant at the manifold $M$ and having a vanishing first derivative with respect to $s$ for $s=0$. Notice that due to the dimension of $\Lambda$,  $v(s,x^{i})$ does not necessarily have dimension of energy. In this section, additional considerations on the definition of the thin-shell limit 
will be analyzed by relating it to the geometrical distortion of the shell.

In order to study the vicinity of the manifold $M$, let
us start with a Taylor expansion along the orthogonal direction
\begin{eqnarray}
v(s,x^{i}) & =K+\frac{1}{2!}s^{2}\frac{\partial^{2}v(s,x^{i})}{\partial s^{2}}\Bigg|_{\textrm{M}}+O(s^{3}),\label{eq:expansao}
\end{eqnarray}
with $K=v|_\textrm{M}$ being a constant at the minimum
$M$, which is characterized by $s=0$, and the first derivative in $s$ vanishes there.  
The second derivative of the potential, in the case of the bubble trap, defines the harmonic frequency $\Lambda \Omega$ around the vicinity of the shell. Such derivative follows from
\begin{eqnarray}
\frac{\partial^{2}v(s,x^{i})}{\partial s^{2}}\Bigg|_{\textrm{M}}= \sum_{\alpha , \beta =1}^{3}\frac{\partial^{2}v(x,y,z)}{\partial r^{\alpha}\partial r^{\beta}}\Bigg|_{\textrm{M}}n^{\alpha}n^{\beta}=\hat{n}\cdot(\nabla\nabla v(x,y,z)|_{\textrm{M}})\cdot\hat{n}=m \Omega^{2}(x,y,z),\label{secondderivative} 
\end{eqnarray}
where $m$ defines the mass of the particles,  $r^{1}=x$, $r^{2}=y$, $r^{3}=z$ represent the Cartesian coordinates, and $n^{\alpha}$, $n^{\beta}$ denote the respective components of the normal vector. Observe that $\Omega$ does not have the same dimension as the harmonic frequency $\Lambda\Omega$, due to the prefactor $\Lambda$.

In order to take into account the geometry in the usual bubble trap experiments \cite{bubble1, bubble2, bubble3, bubble4, bubble5, bubble6, bubble7, bubble8, CAL1, CAL2, CAL3, CAL4, CAL5, CAL6, CAL7}, let us consider the finite factor of our potential in a generalized form as
\begin{eqnarray}
v(x,y,z)\equiv f(x^{2}+y^{2}+(1+\epsilon)z^{2}).
\label{funcaof}
\end{eqnarray}
In this case the geometry of the potential is well established to be an ellipsoidal surface,
which becomes spherical for $\epsilon=0$. Thus, our manifold $M$ is characterized by $x^{2}+y^{2}+(1+\epsilon)z^{2}=A^{2}$, so we conclude $f(x^{2}+y^{2}+(1+\epsilon)z^{2})|_{\textrm{M}}=f(A^2)=\textrm{const.}$ and we have $f'(A^2)=0$.

With such a general form, let us explicitly calculate $m\Omega^{2}(s,x^i)$.  The derivatives can be written in vector form as $\nabla v(x,y,z)|_{\textrm{M}}\cdot\hat{n} 
=f'(x^{2}+y^{2}+(1+\epsilon)z^{2})[2x\hat{x}+2y\hat{y}+2(1+\epsilon)z\hat{z}]|_{\textrm{M}}\cdot\hat{n}$. With $\hat{n}$ expressed in Cartesian coordinates, we obtain $\nabla v(x,y,z)|_{\textrm{M}}\cdot\hat{n} = 2f'(A^{2})\sqrt{x^{2}+y^{2}+(1+\epsilon)^{2}z^{2}}=0$.
By working out the second derivatives as well, we read off from (\ref{secondderivative}) that $m\Omega\textrm{\texttwosuperior}(x,y,z)=4f''(A^{2})[A^{2}+\epsilon(1+\epsilon)z^{2}]$, which reduces in terms of the GNCS to
\begin{eqnarray}
m\Omega\textrm{\texttwosuperior}(x^i) & =4f''(A^{2})A^{2}(1+\epsilon\cos^{2}\nu).\label{frequency}
\end{eqnarray}
This gives the final form for the harmonic frequency $\Lambda\Omega$ considering our generalized potential in form of an ellipsoidal surface with the appropriate constraints. Therefore, the confinement strength turns out to be proportional to the geometrical distortion of the ellipsoid through the dependence on the parameter $\epsilon$.
Moreover, the dependence on the angle $\nu$ establishes that the confinement varies from the poles to the equator of the ellipsoid. Apart from the mentioned conditions, the function $f$ can be quite general, only with the natural assumption that $f''(A^2)>0$.

Now, some careful considerations must be made for the thin-shell limit.
For infinitely tight potentials, we know in advance, that the motion of particles along the normal direction is restricted  to the ground state of an harmonic oscillator with frequency $\Lambda\Omega(x^i)$ which is described by a Gaussian wave function with width $\alpha=\sqrt{\hbar/m\Lambda\Omega}$. In previous works, mainly devoted to the spherical case, some authors defined the thin-shell limit as a rather generic situation, where the radius $R$ of the sphere-shaped trap is much larger than the thickness $\alpha$ of the shell. Since these are not the only length scales in this system, different results can be obtained by either considering $R \rightarrow\infty$ or $\alpha \rightarrow 0$. For example in  \cite{vortex8} the authors considered the situation where $R \rightarrow\infty$. In the numerical work \cite{bubbleteo4}, the authors made some qualitative comparisons between $\alpha$ and $R$. In some papers \cite{bubbleteo8, bubbleteo6, bubbleteo7}, the ratio $R/\alpha$ is chosen to be as large as possible without more detailed considerations. The objective in this paper is to consider the specific case $\alpha \rightarrow 0$, i.e., the case where $\alpha$ is much smaller than any other length scale of the system. This constitutes a more precise definition of a thin-shell limit, which is equivalent to consider $\Lambda \rightarrow\infty$.




As already mentioned, in the thin-shell limit we expect the motion in the normal direction to be confined to the ground state of a harmonic oscillator with frequency $\Lambda \Omega$. Since for the ellipsoidal case, $\Omega$ is space dependent, every particle in the system will experience a site-dependent energy due to its confinement. Such an energy is provided by the ground-state energy of a harmonic oscillator with frequency $\Lambda \Omega$, i.e.,
\begin{eqnarray}
E=\frac{\hbar\Lambda\Omega}{2}=\frac{\hbar\Lambda}{2}\sqrt{\frac{4f''(A^{2})A^{2}(1+\epsilon\cos^{2}\nu)}{m}}.
\label{E1}
\end{eqnarray}
Here we see that for $\epsilon= 0$, this energy diverges for $\Lambda\rightarrow\infty$, thus adding a uniform infinite potential which does not have any consequence for the dynamics of the particles moving in the manifold. However for $\epsilon \neq 0$ the situation changes considerably since the particles would be subjected to an infinite space-dependent potential. Let us for example consider the energy difference between the poles and the equator $\Delta E =E(\nu=0)-E(\nu=\pi/2)$, for which we find $\Delta E \propto \Lambda$ for any non-vanishing $\epsilon$, which means that it diverges in the thin-shell limit. This would then induce a
complete localization of particles either at the poles or at the equator, depending on the sign of $\epsilon$.
Thus, in order to restrict our limit to physically acceptable situations, we must consider only the case of small geometrical distortions
\begin{eqnarray}
\epsilon=\Lambda^{-1}\bar{\epsilon},
\label{geometrical}
\end{eqnarray}
where $\bar{\epsilon}$ is considered to be finite.
By taking into account (\ref{geometrical}), the induced energy difference then becomes
\begin{eqnarray}
\Delta E =\frac{\hbar}{4}\sqrt{\frac{4f''(A^{2})A^{2}}{m}}\bar{\epsilon} + O(\Lambda^{-1}),\label{eq:potcomp}
\end{eqnarray}
which is finite in the thin-shell limit.
For arbitrary values of $\nu$, Eq.~(\ref{E1}) becomes
\begin{eqnarray}
E=\frac{\hbar\Lambda}{2}\sqrt{\frac{4f''(A^{2})A^{2}}{m}} + \bar{\epsilon}\frac{\hbar}{4}\sqrt{\frac{4f''(A^{2})A^{2}}{m}}\cos^{2}\nu + O(\Lambda^{-1}).
\label{E2}
\end{eqnarray}
Thus, we recognize that the  energy per particle generated due to the compactfication process separates into one physically irrelevant infinite site-independent part and a finite site-dependent part. It means that, in the thin-shell limit, each particle is effectively subjected to a compactification potential given by
\begin{eqnarray}
V_{\rm comp}= \bar{\epsilon}\frac{\hbar}{4}\sqrt{\frac{4f''(A^{2})A^{2}}{m}}\cos^{2}\nu .
\label{Vcomp}
\end{eqnarray}
In experimental terms, the infinitesimal eccentricity $\epsilon = \Lambda^{-1} \bar{\epsilon}$ corresponds to the situation, where the difference between the larger axis and the smaller axis of the ellipsoid is of the same order of the Gaussian width $\alpha$. This means that, in order to avoid that the particles collapse to either the poles or the equator, the ellipsoidal eccentricity must be kept within such bounds.  

In the next subsections, let us consider an experimental realization of the general potential
(\ref{funcaof}) as a generic example where the theory developed in here can be realized.
\subsection{Confinement potential in experiments}
In this subsection, we apply the calculations to a particular confinement potential. Note that there are some variations of this potential in the literature
that differ in the way the potential is defined, for instance, by including gravity effects or considering frequency anisotropy \cite{bubble1, bubble2}.

Let us consider here the more
concrete experimental example in \cite{bubble3} as an application of our theory, where
\begin{eqnarray}
V_{\textrm{E}}(x,y,z)=\sqrt{\left[V_{\textrm{trap}}(x,y,z)-\hbar\triangle\right]^{2}+(\hbar\Omega_{\textrm{RF}})^{2}},
\end{eqnarray}
with $V_{\textrm{trap}}(x,y,z)=\omega^{2}v{}_{\textrm{trap}}(x,y,z)= m\omega^{2}[x^{2}+y^{2}+(1+\epsilon)z^{2}]/2$. Here $\hbar\triangle=\hbar\omega_{\textrm{rf}}-V_{0}$ denotes the rf
detuning with respect to the rf transition at the center of the magnetic trap, and $\Omega_{\textrm{RF}}=g \mu_{\rm B}B_{1}/2\hbar$ stands for the Rabi frequency of the magnetic field $B_{1}\cos(\omega_{\textrm{rf}}t)$ with the Land\'e factor $g=1/2$ and the Bohr magneton $\mu_{\textrm{B}}$, whereas $\epsilon$ represents the geometrical distortion of the ellipsoid. This potential has a local minimum provided that $V_{\textrm{trap}}(x,y,z)=\hbar\triangle$ for the state $F=2$, $m_{\textrm{F}}=2$. 

Let us express the potential according to Eq. (\ref{genpot}), i.e.,
\begin{eqnarray}
V_{\textrm{E}}(x,y,z)=\omega^{2}\sqrt{\left[v{}_{\textrm{trap}}(x,y,z)-\frac{\hbar\triangle}{\omega^{2}}\right]^{2}+\Bigg(\frac{\hbar\Omega_{\textrm{RF}}}{\omega^{2}}\Bigg)^{2}}=\omega^{2}v_{\textrm{E}}(x,y,z),
\end{eqnarray}
which defines a function $f$ according to (\ref{funcaof}), where $\Lambda^{2}$ equals to $\omega^{2}$ in this particular case. 
The thin-shell limit for such a potential can be obtained by considering $\omega\rightarrow\infty$, $\triangle\rightarrow\infty$, and $\Omega_{\small \rm RF}\rightarrow\infty$, while the radii $\hbar\triangle/\omega^{2}$ and $\hbar\Omega_{\textrm{RF}}/\omega^{2}$ are kept finite. Also $\epsilon\omega$ must be kept finite in order to prevent the collapse of the condensate.
The expression for the harmonic frequency $\Lambda\Omega$ according to (\ref{frequency}) and considering the GNCS with $A=\sqrt{2\hbar\triangle/m\omega^{2}}$ reads
\begin{eqnarray}
m\Omega_{\textrm{E}}^{2}(x^{i}) & =\frac{2m\triangle}{\Omega_{\textrm{RF}}}(1+\epsilon\cos^{2}\nu).
\label{bubble-trap}
\end{eqnarray}
In more experimental terms, the range of parameters corresponding to the thin-shell limit occur for $\alpha \ll A$. Thus, the thin-shell limit corresponds to the inequality
\begin{eqnarray}
\frac{2\Delta}{\omega}\gg\sqrt{\frac{\Omega_{\textrm{RF}}}{2\Delta}},
\label{inequality}
\end{eqnarray}
which represents a condition involving all three frequencies $\Delta$, $\omega$, $\Omega_{\textrm{RF}}$ of the bubble trap potential (\ref{bubble-trap}).
Conversely, in order to prevent the collapse of the condensate, we must also have $\epsilon A \sim \alpha$. This leads to the eccentricity
\begin{eqnarray}
\epsilon \sim \sqrt{\frac{\omega}{2\Delta}\sqrt{\frac{\Omega_{\textrm{RF}}}{2\Delta}}},
\end{eqnarray}
which is small due to the inequality (\ref{inequality}).
Within the thin-shell limit for this particular experimental case Eq.~(\ref{genpot}) with the aid of (\ref{eq:expansao}) becomes
\begin{eqnarray}
V_{\textrm{E}}(s,x^{i}) =\omega^{2}\left[\frac{\hbar\Omega_{\textrm{RF}}}{\omega^{2}}+\frac{1}{2!}s^{2}\frac{2m\triangle}{\Omega_{\textrm{RF}}}(1+\omega^{-1}\bar{\epsilon}\cos^{2}\nu)+O(s^{3})\right].
\end{eqnarray}
And the corresponding compactification potential follows from Eq. (\ref{Vcomp})
\begin{eqnarray}
V_{\textrm{comp}}= \bar{\epsilon}\frac{\hbar}{4}\sqrt{\frac{2\triangle}{\Omega_{\textrm{RF}}}}\cos^{2}\nu .
\end{eqnarray} 
Such expressions confirm that the general theory presented here is well suited to deal with the already created experimental bubble-trap potentials.

In order to provide a concrete example, one can consider $^{87}$Rb atoms and the frequencies $\omega = 2\pi \times 173$\,Hz, $\Delta  =
2\pi \times 30$\,kHz, and $\Omega_{\textrm{RF}} = 2\pi \times 3$\,kHz \cite{CAL5}, which amounts to $\alpha = 0.4\, \mu\mbox{m}$ and
$A = 15\,\mu\mbox{m}$. This fulfills, indeed, the thin-shell limit condition $\alpha \ll A$ and prevents the collapse for the quite small eccentricity $\epsilon \sim 0.027$.
\section{Effective Hamiltonian in the thin-shell limit}
In this section, we will consider the thin-shell limit for interacting particles. Let us begin with the usual 3D Gross-Pitaevskii Hamiltonian
\begin{eqnarray}
H & =\int d^{3}x \, \bigg\{\frac{\hbar\textrm{\texttwosuperior}}{2m} |\nabla \Psi |^2  +V(x,y,z)|\Psi |^2+\frac{g_{\textrm{int}}}{2} |\Psi |^4 \bigg\},
\label{H1}
\end{eqnarray}
written in Cartesian coordinates, where $V$ stands for the overall potential and $g_{\rm int}$ represents the interaction parameter. The total number of particles is given by
\begin{eqnarray}
N & =\int d^{3}x\; \Psi^*\Psi.
\label{N1}
\end{eqnarray}
Rewriting the Hamiltonian  (\ref{H1}) in the GNCS by using the Laplace-Beltrami form of the Laplace operator yields approximately 
\begin{eqnarray}
H & \approx \int dx^i \int_{s\textrm{\textsuperscript{-}}(x^{i})}^{s\textrm{\textsuperscript{+}}(x^{i})}ds\sqrt{|g|}\bigg\{-\frac{\hbar\textrm{\texttwosuperior}}{2m}\Psi^{*}\bigg[|g|^{-1/2}\frac{\partial}{\partial s}\bigg(|g|^{1/2}\frac{\partial\Psi}{\partial s}\bigg)+|g|^{-1/2}\frac{\partial}{\partial x^{i}}\bigg(|g|^{1/2}g^{ij}\frac{\partial\Psi}{\partial x^{j}}\bigg)\bigg]\nonumber \\
 & +V(s,x^{i})\Psi^{*}\Psi+\frac{g_{\textrm{int}}}{2}\Psi^{*2}\Psi^{2}\bigg\} \label{H2}
\end{eqnarray}
with the abbreviation
$\int dx^i=\int dx^1 \int dx^2$. 
Here the reduced covariant 2D metric turns out to be
 \begin{eqnarray}
g_{ij}=
\left[
\begin{array}{ccc}
\Bigg(\cos^{2}\nu+\frac{\sin^{2}\nu}{1+\epsilon}\Bigg)\Bigg(A+s\frac{1+\epsilon}{(1+\epsilon\cos^{2}\nu)^{3/2}}\Bigg)^{2} & 0 \\
0 & \sin^{2}\nu\Bigg(A+s\frac{1}{\sqrt{1+\epsilon\cos^{2}\nu}}\Bigg)^{2} 
\end{array} 
\right],
\label{r_metric}
\end{eqnarray} 
which has only the elements corresponding to the tangent coordinates. The limits of the $s$ integral are chosen in such a way that both $(\Lambda\Omega m s\textrm{\textsuperscript{-}}^2)/\hbar \gg 1$ and $(\Lambda\Omega m s\textrm{\textsuperscript{+}}^2)/\hbar \gg 1$. This assures that the difference between (\ref{H1}) and (\ref{H2}) decays exponentially for $\Lambda\rightarrow\infty$ since  $\Psi \sim e^{-[\Lambda\Omega(x^{i}) m s^2]/2\hbar}$ in the thin-shell limit.

In order to simplify our calculations, we follow \cite{bubbleteo9} and consider the alternative wave function $\widetilde{\Psi}$ 
\begin{eqnarray}
\Psi(s,x^{i})=\frac{|g_{0}(x^{i},\epsilon)|^{1/4}}{|g(s,x^{i},\epsilon)|^{1/4}}\widetilde{\Psi}(s,x^{i}),
\end{eqnarray}
where $|g_0(x^i,\epsilon)|=|g(0,x^i,\epsilon)|$. Note that the normalization condition \ref{N1} for $\tilde{\Psi}$ then reduces to
\begin{eqnarray}
N & \approx\int dx^i \sqrt{|g_{0}|}\int_{s\textrm{\textsuperscript{-}}(x^{i})}^{s\textrm{\textsuperscript{+}}(x^{i})}ds \tilde{\Psi}^*\tilde{\Psi}.
\end{eqnarray}
This allows us to use the 2D $s$-independent Jacobian $\sqrt{|g_0|}$ also for the Hamiltonian
%
%
\begin{eqnarray}
&&H  \approx \int dx^i  \sqrt{|g_{0}|}\int_{s\textrm{\textsuperscript{-}}(x^{i})}^{s\textrm{\textsuperscript{+}}(x^{i})}ds\bigg\{-\frac{\hbar\textrm{\texttwosuperior}}{2m}\gamma^{-1/4}\widetilde{\Psi}^{*}\frac{\partial}{\partial s}\left[\gamma^{1/2}\frac{\partial}{\partial s}\bigg(\gamma^{-1/4}\widetilde{\Psi}\bigg)\right] \label{HH}\\
& & -\frac{\hbar\textrm{\texttwosuperior}}{2m}|g_{0}|^{-1/2}\gamma^{-1/4}\widetilde{\Psi}^{*}\frac{\partial}{\partial x^{i}}\left[|g_{0}|^{1/2}\gamma^{1/2}g^{ij}\frac{\partial}{\partial x^{j}}\bigg(\gamma^{-1/4}\widetilde{\Psi}\bigg)\right] +V(s,x^{i},\epsilon)\widetilde{\Psi}^{*}\widetilde{\Psi}+\frac{g_{\textrm{int}}}{2}\gamma^{-1/2}\widetilde{\Psi}^{*2}\widetilde{\Psi}^{2}\bigg\},\nonumber
\end{eqnarray}
where we have introduced  $\gamma(s,x^i,\epsilon)=|g(s,x^i,\epsilon)|/|g_{0}(x^i,\epsilon)|$.
%
%
As already mentioned, in the thin-shell limit we expect $\widetilde{\Psi}\sim e^{-(\Lambda\Omega m s^2)/2\hbar}$, i.e., $\widetilde{\Psi}$ depends implicitly on $\Lambda$. In order to make such a dependency explicit, while maintaining the normalization of $\widetilde{\Psi}$ as well as to keep the interaction term finite in the limit $\Lambda\rightarrow\infty$, we perform the following scale transformations  
\begin{eqnarray}
\begin{cases}
s=\Lambda^{-1/2}u,\\
\widetilde{\Psi}=\Lambda^{1/4}\psi  ; \quad\widetilde{\Psi}^{*}=\Lambda^{1/4}\psi^{*},\\
g_{\textrm{int}}=\Lambda^{-1/2}\bar{g}_{\textrm{int}},\\
\end{cases}\label{rescale}
\end{eqnarray}
that gives us some extra control over the Taylor expansions to be performed inside the integrals. These rescaled quantities are the normal direction $s$, the wave function $\psi$, and the particle interaction $g_{\textrm{int}}$.

Introducing the abbreviations $\gamma_1 = \partial\gamma/\partial s$, $\gamma_2 = \partial^2\gamma/\partial s^2$,
we obtain together
with (\ref{genpot}) and (\ref{rescale}) for the Hamiltonian (\ref{HH})
\begin{eqnarray}
H & \approx \int dx^i \sqrt{|g_{0}|}\int_{\Lambda^{1/2}s\textrm{\textsuperscript{-}}(x^{i})}^{\Lambda^{1/2}s\textrm{\textsuperscript{+}}(x^{i})}du\bigg\{-\Lambda\frac{\hbar\textrm{\texttwosuperior}}{2m}\psi^{*}\frac{\partial^{2}\psi}{\partial u^{2}}\nonumber -\frac{\hbar\textrm{\texttwosuperior}}{2m}\left(\frac{3}{16}\frac{\gamma_{1}^{2}}{\gamma^{2}}-\frac{1}{4}\frac{\gamma_{2}}{\gamma}\right)\psi^{*}\psi\nonumber \\
 & -\frac{\hbar\textrm{\texttwosuperior}}{2m}\gamma^{-1/4}\psi^{*}|g_{0}|^{-1/2}\frac{\partial}{\partial x^{i}}\left[(|g_{0}|^{1/2}\gamma^{1/2}g^{ij}\frac{\partial}{\partial x^{j}}\bigg(\gamma^{-1/4}\psi\bigg)\right]\nonumber \\
 & +\Lambda\textrm{\texttwosuperior}v(\Lambda^{-1/2}u,x^{i},\Lambda^{-1}\epsilon)\psi^{*}\psi+\frac{\bar{g}_{\textrm{int}}}{2}\gamma^{-1/2}\psi^{*2}\psi^{2}\bigg\}.
\end{eqnarray}
Now we can expand both $v$ and $\gamma$ in a power series with respect to $\Lambda^{-1/2}$ which yields
\begin{eqnarray}\hspace*{-1cm}
\Lambda^{2}v(\Lambda^{-1/2}u, x^i,\Lambda^{-1}\bar\epsilon) &=&\Lambda^{2}K+\Lambda\frac{1}{2!}4f''(A^2)A^2 u^{2}\nonumber \\
&&+\Lambda^{1/2}\frac{1}{3!}\left[8f'''(A^{2})A^{3}+12f''(A^{2})A\right]u{}^{3}+\frac{\bar{\epsilon}}{2!}4f^{\prime\prime}(A^{2})A^{2}\cos^{2}\nu\, u^{2}   \nonumber \\
&&+\frac{1}{4!}\left[16f^{IV}(A^{2})A{}^{4}+48f'''(A^{2})A^{2}+12f''(A^{2})\right]u{}^{4}+O(\Lambda^{-1/2}).\label{pottaylor}
\end{eqnarray}
Note that we have
\begin{eqnarray}
\gamma= 1+ O(\Lambda^{-1/2}),
\end{eqnarray}
and according to \ref{apendice} we conclude that
\begin{eqnarray}
\left(\frac{3}{16}\frac{\gamma_{1}^{2}}{\gamma^{2}}-\frac{1}{4}\frac{\gamma_{2}}{\gamma}\right)=O(\Lambda^{-1/2}).
\end{eqnarray}
Note that the presence of the harmonic frequency in (\ref{frequency}) considering (\ref{geometrical}) leads to $m\Omega^2 = 4f''(A^2)A^2 + \Lambda ^{-1}\bar{\epsilon}4f''(A^2)A^2\cos^2\nu = m\Omega^2_0 + \Lambda^{-1}\bar{\epsilon} m\Omega^2_0 \cos^2\nu$, where
\begin{eqnarray}
m\Omega^2_0 = 4f''(A^2)A^2.
\end{eqnarray}
Substituting these equations and neglecting exponentially decaying contributions, the Hamiltonian can be written as $H=H_{2}+H_{1}+H_{1/2}+H_{0}+\textrm{O}(\Lambda^{-1/2})$, where the respective terms are sorted according to decreasing contributions with
\begin{eqnarray}
H_{2}  &=&\Lambda^{2}K\int dx^i \int du\sqrt{|g_{0}|}\psi^{*}\psi=\Lambda^{2}KN,\\
H_{1}  &=&\Lambda\int dx^i \int du\sqrt{|g_{0}|}\left( -\frac{\hbar\textrm{\texttwosuperior}}{2m}\psi^{*}\frac{\partial^{2}\psi}{\partial u^{2}}+\frac{1}{2}m\Omega^2_{0}u^{2}\psi^{*}\psi\right) ,\\
H_{1/2}  &=&\Lambda^{1/2}\int dx^i \int du\sqrt{|g_{0}|}\left[ \frac{4}{3}f'''(A^{2})A^{3}+2f''(A^{2})A\right]u{}^{3}\psi^{*}\psi ,\\
H_{0} & =&\int dx^i\int du\sqrt{|g_{0}|}\bigg\{-\frac{\hbar\textrm{\texttwosuperior}}{2m}\psi^{*}|g_{0}|^{-1/2}\frac{\partial}{\partial x^{i}}\bigg(|g_{0}|^{1/2}g^{ij}_0\frac{\partial}{\partial x^{j}}\psi\bigg)
+\frac{\bar{\epsilon}}{2}m\Omega^2_{0}u^{2}\cos^2\nu\;\psi^{*}\psi
\nonumber \\
 & &+ \left[\frac{2}{3}f^{IV}(A^{2})A{}^{4}+2f'''(A^{2})A^{2}+\frac{1}{2}f''(A^{2})\right]u{}^{4}\psi^{*}\psi+\frac{\bar{g}_{\textrm{int}}}{2}\psi^{*2}\psi^{2}\Bigg\},\label{eq:H0}
\end{eqnarray}
with limits for the integrals in $u$ being $-\infty$ to $\infty$. Since $H_{2}$ turns out to be a simple additive constant, our analysis effectively starts with $H_{1}$. 

Note that $H_{1}$ lacks an interaction term and essentially corresponds to the harmonic oscillator Hamiltonian whose spectrum is given by the linear eigenvalue equation
\begin{eqnarray}
-\frac{\hbar\textrm{\texttwosuperior}}{2m}\frac{\partial^{2}\psi}{\partial u^{2}}+\frac{1}{2}m\Omega_{0}^2u^{2}\psi=\frac{E_{1}}{\Lambda}\psi.\label{HOequation}
\end{eqnarray}
Therefore, at order $\Lambda^{0}$ the degenerate eigenfunctions of \ref{HOequation} are given by
\begin{eqnarray}
\psi_{n,l}^{(0)}(u,x^i)&=&\psi_{n\perp}(u)\xi_l(x^{i})\nonumber\\ &=&\frac{1}{\sqrt{2^{n}n!}}\left(\frac{m\Omega_{0}}{\pi\hbar}\right)^{1/4}\exp\left(-\frac{m\Omega_{0}}{2\hbar}u^{2}\right)H_{n}\left(\sqrt{\frac{m\Omega_{0}}{\hbar}}u\right)\xi_l(x^i),\label{psi}
\end{eqnarray}
where $H_{n}$ denote the Hermite polynomials, $\psi_{n\perp}$ represent the harmonic oscillator eigenstates, and $\xi_l$  stand for arbitrary functions of $x^i$ obeying the condition $\sum_l\int dx^1 \int x^2\sqrt{|g_{0}|}\xi^{\ast}_l(x^{i})\xi_l(x^{i}) = N$. The dominant contribution to the energy spectrum of system reads
\begin{eqnarray}
E_n^{(1)}= N \Lambda \hbar \Omega_0\left(\frac{1}{2}+n\right).
\label{eq:Spectrum-full}
\end{eqnarray}
Therefore, in the thin-shell limit, where we have $\Lambda\rightarrow\infty$, the system acquires infinitely separated energy bands. This implies that, physically, only the lowest energy band with energy $E_1=E_0^{(1)}= N \Lambda \hbar \Omega_0/2$  will be populated with particles, i.e., $\xi_l=0$ for $l\neq 0$. In particular, the normalization condition for $\xi_0$ becomes
\begin{eqnarray}
\int dx^i \sqrt{|g_{0}|}\xi^{\ast}_0(x^{i})\xi_0(x^{i})= N.
\label{constraint}
\end{eqnarray}

Now, let us deal with the effects of the $H_{1/2}$ contribution to the total Hamiltonian. Similar to $H_1$, also $H_{1/2}$ acts only in the subspace defined by the variable $u$. This means that it will induce corrections of order $\Lambda^{-1/2}$ to $\psi_{n\perp}$ as well as corrections of order $\Lambda^{1/2}$ to the energies (\ref{eq:Spectrum-full}). Such corrections of order $\Lambda^{-1/2}$ to $\psi_{n\perp}$ can be neglected in the thin-shell limit, while the contribution with power $\Lambda^{1/2}$ to (\ref{eq:Spectrum-full}) vanishes since it involves the infinite integral of an odd function of $u$ according to
\begin{eqnarray}\hspace*{-0.8cm}
E_{n}^{(1/2)} =\Lambda^{1/2}\int dx^i \int^{+\infty}_{-\infty} du\sqrt{|g_{0}|}\left[\frac{4}{3} f'''(A^{2})A^{3}+2f''(A^{2})A \right]u^{3}\psi^{*(0)}_{n,0}\psi^{(0)}_{n,0}=0.
\end{eqnarray}
Although the contribution to the energy with power $\Lambda^{1/2}$ vanishes, an application of second-order perturbation theory considering $H_{1/2}$ as a perturbation over $H_1$ shows that contributions with power $\Lambda^0$ to (\ref{eq:Spectrum-full}) do not vanish. In particular the correction $E^{(0)}_0$ to $E^{(1)}_0$ due to $H_{1/2}$ is
\begin{eqnarray}
E_{0}^{(0)} &=-\frac{11 N \hbar^2}{18 m} \Bigg(\frac{2f'''(A^{2})A^{3}+3f''(A^{2})A}{4f''(A^2)A^2}\Bigg)^{2}.
\label{H1/2}
\end{eqnarray}
Further contributions to (\ref{eq:Spectrum-full}) due to $H_{1/2}$ are of order $\Lambda^{-1/2}$ and are negligible in the thin-shell limit.

Finally, we must consider the effect of $H_0$ for the energy spectrum and eigenstates. The first thing to observe is that $H_0$ also acts over the subspace defined by the coordinates $x^i$, which implies that $H_0$ is able to break the degeneracy. Therefore, at order $\Lambda^0$ the system is effectively separated into disjoint subspaces, each one having wave functions given by $\psi_{n\perp}(u)\xi_n(x^i)$. According to (\ref{eq:Spectrum-full}), the energy gap between such subspaces is of the order of $\Lambda$, which means that in the thin-shell limit $\Lambda\rightarrow\infty$ only the subspace with lowest energy becomes occupied. Consequently, the system wave function in the thin-shell limit must be
\begin{eqnarray}
\psi(u,x^{i})=\left(\frac{m\Omega_{0}}{\pi\hbar}\right)^{1/4}\exp\left(-\frac{m\Omega_{0}}{2\hbar}u^{2}\right)\xi(x^{i}),\label{eq:degenerate-ground}
\end{eqnarray}
where $\xi(x^{i})=\xi_0(x^{i})$. Substituting (\ref{eq:degenerate-ground}) into $H_0$ and integrating where possible yields the constant contribution
\begin{eqnarray}
 C_0&=&\frac{N}{8}\frac{\hbar^{2}}{m}\,\frac{4f^{IV}(A^{2})A^{4}+12f'''(A^{2})A^{2}+3f''(A^{2})}{4f''(A^2)A^2}
\end{eqnarray}
in addition to the effective Hamiltonian
\begin{eqnarray}
H_{\textrm{eff}} =  \int dx^i \sqrt{|g_{0}|}\bigg\{-\frac{\hbar\textrm{\texttwosuperior}}{2m}\xi^{*}|g_{0}|^{-1/2}\frac{\partial}{\partial x^{i}}\bigg(|g_{0}|^{1/2}g^{ij}_0\frac{\partial}{\partial x^{j}}\xi\bigg) +
\frac{\bar{\epsilon}}{4}\hbar \Omega_{0}\cos^{2}\nu\;\xi^{*}\xi +\frac{\bar{g}_{2D}}{2}\xi^{*2}\xi^{2}\Bigg\}.
\end{eqnarray} 
The latter simplifies to
\begin{eqnarray}
H_{\textrm{eff}} & = & \int_{0}^{\pi}d\nu\int_{0}^{2\pi}d\phi A^{2}\sin\nu\bigg\{-\frac{\hbar\textrm{\texttwosuperior}}{2mA^{2}\sin\nu}\xi^{*}\left[\frac{\partial}{\partial\nu}\left(\sin\nu\frac{\partial\xi}{\partial\nu}\right)+\frac{\partial}{\partial\phi}\left(\frac{1}{\sin\nu}\frac{\partial\xi}{\partial\phi}\right)\right]\nonumber \\
&  & +\frac{\bar{\epsilon}}{4}\hbar \Omega_{0}\cos^{2}\nu\;\xi^{*}\xi 
+\frac{\bar{g}_{2D}}{2}\xi^{*2}\xi^{2}\Bigg\},
 \label{Heff}
\end{eqnarray}
where the resulting effective interaction strength in 2D turns out to be
\begin{eqnarray}
\bar{g}_{\rm 2D}=\left(\frac{m\Omega_{0}}{2\pi\hbar}\right)^{1/2}\bar{g}_{\textrm{int}}=\left(\frac{m\Omega_{0}\Lambda}{2\pi\hbar}\right)^{1/2}g_{\textrm{int}}.
\end{eqnarray}
With this, we conclude that (\ref{Heff}) is the appropriate Hamiltonian in the thin-shell limit apart from the diverging additive constant $E_2+E^{(1)}_0+E^{(0)}_0+C_0$.
\section{Particular cases}
Evaluating the functional derivative of the effective Hamiltonian (\ref{Heff}), it is possible to obtain the wave function that minimizes it. To this end, however, we have to take into account the constraint (\ref{constraint}) and its 
corresponding Lagrange multiplier, the chemical potential $\mu$ according to
\begin{eqnarray}
\frac{\delta H_{\textrm{eff}}}{\delta\xi^{\ast}(x^{i})}-\mu\frac{\delta N}{\delta\xi^{\ast}(x^{i})}=0,
\end{eqnarray}
which finally leads to
\begin{eqnarray}
 \hspace*{-1.8cm} -\frac{\hbar\textrm{\texttwosuperior}}{2mA^{2}\sin\nu}\left[\frac{\partial}{\partial\nu}\left(\sin\nu\frac{\partial\xi}{\partial\nu}\right)+\frac{\partial}{\partial\phi}\left(\frac{1}{\sin\nu}\frac{\partial\xi}{\partial\phi}\right)\right]
 +\frac{\bar{\epsilon}}{4}\hbar \Omega_{0}\cos^{2}\nu\xi
 +\bar{g}_{2D}\xi^{*}\xi^{2}=\mu\xi.
 \label{casca}
\end{eqnarray}
Let us consider now some particular cases for this 2D time-independent GPE.
\subsection{Spherical shell}
At first we deal with the particular situation of the spherical shell, i.e.~$\bar{\epsilon}=0$, which allows some simplifications. Most importantly, the wave function corresponding to the lowest energy state $\xi(x^{i})$ is a constant along the shell, since the BEC particles are uniformly distributed around the bubble.
Let us determine the constant value of $\xi$ through the normalization (\ref{constraint}), which yields for the ground state of the system
\begin{eqnarray}
\xi=\sqrt{\frac{N}{A^{2}4\pi}}.
\end{eqnarray}
Substituting this value for the wave function into the effective Hamiltonian (\ref{Heff}), we find with $\bar{\epsilon}=0$ the corresponding energy
\begin{eqnarray}
E_{\textrm{S}} = \frac{\bar{g}_{2D}}{8A^{2}\pi}N^{2}.
\end{eqnarray}
This is the $\Lambda^{0}$ contribution to the energy for the spherical shell. Thus, the total ground-state energy amounts in this case to $E_2+E^{(1)}_0+E^{(0)}_0+C_0+E_\textrm{S}$, which is consistent with Ref. \cite{bubbleteo1}.
\subsection{Thomas-Fermi approximation}
For situations, where the interaction energy is much larger than the kinetic energy, the so-called Thomas-Fermi approximation can be applied. In case of the effective Hamiltonian \ref{Heff}, this corresponds to the situation where $N\rightarrow \infty$ or $A\rightarrow \infty$.
In order to calculate the wave function $\xi(x^{i})$, we take  (\ref{casca}) without the kinetic energy term, which gives
\begin{eqnarray}
\xi=\sqrt{\frac{\mu-\frac{\bar{\epsilon}}{4}\hbar \Omega_{0}\cos^{2}\nu}{\bar{g}_{2D}}}.
\end{eqnarray}
As expected, $\xi(x^{i})$ has an angular dependence along the shell. Inserting it into equation (\ref{Heff}) and solving the integral, we find the Thomas-Fermi energy 
\begin{eqnarray}
E_{\textrm{TF}} =\frac{ \pi A^{2} (80\mu ^{2}- \bar{\epsilon}^{2} \hbar ^{2} \Omega_{0}^{2} )}{40 \bar{g}_{2D}},
\end{eqnarray}
which represents the ground state $\Lambda^{0}$  energy contribution without the diverging additive constants. Taking into account (\ref{constraint}), it can be rewritten in terms of the total number of particles $N$ as
\begin{eqnarray}
E_{\textrm{TF}} =\frac{\pi A^{2}}{40 \bar{g}_{2D}}\Bigg(\frac{5 \bar{g}_{2D}^{2} N^{2}}{\pi ^{2} A^{4}}+\frac{10}{3 \pi A^{2}}\bar{g}_{2D} N \bar{\epsilon}  \hbar \Omega_{0}-\frac{4 \bar{\epsilon}^{2} \hbar ^{2} \Omega_{0}^{2}}{9}\Bigg).
\end{eqnarray}
Thus, the total ground-state energy in this case is $E_2+E^{(1)}_0+E^{(0)}_0+C_0+E_\textrm{TF}$, which is also consistent with Ref. \cite{bubbleteo1}.
\section{Perturbative solutions}
Although the main result of our work is the step-by-step construction of the theory for BECs in the thin-shell limit of a bubble trap, this paper also provides solutions to particular cases of the spherical shell and the Thomas-Fermi regime for the effective Hamiltonian (\ref{Heff}), thus revealing some mathematical aspects of the corresponding wave functions. Now, in this chapter, we determine both the ground-state wave function and the excitation frequencies perturbatively by considering the  geometrical distortion $\bar{\epsilon}$ as the smallness parameter.
\subsection{Perturbative ground state}
Let us now perturbatively solve (\ref{casca}), which we rewrite as
\begin{eqnarray}
\frac{1}{2mA^{2}}\hat{L}^{2}\xi+\lambda\cos^{2}(\nu)\xi+\bar{g}_{{\rm 2D}}\left|\xi\right|^{2}\xi=\mu\ \xi,\label{eq:GP1}
\end{eqnarray}
where we have introduced the smallness parameter $\lambda=\bar{\epsilon}\hbar\Omega_{0}/4$. In the following, we also have to take into account the normalization (\ref{constraint}), which reads explicitly $\int dS\left|\xi\right|^{2}=A^{2}\int_{0}^{2\pi}d\phi\int_{0}^{\pi}d\nu\ \sin\nu\ \left|\xi\right|^{2}=N$.

The square of the angular momentum operator $\hat{L}^{2}$ has the eigenvalues $\hbar^{2}l(l+1)$ and normalized eigenfunctions given by the spherical harmonics $Y_{l}^{m}(\nu,\phi)$.
Since the set $\{Y_{l}^{m}\}$ represents a complete orthonormal basis over
the spherical domain, the wave function $\xi$ can be expanded in such a basis. But,
as we are looking for the unique lowest-energy state, which
is represented by a $\phi$-independent real function $\xi(\nu)$,
its expansion becomes
\begin{eqnarray}
\xi(\nu)=\frac{\sqrt{N}}{A}\sum_{i}c_{i}Y_{i}^{0}(\nu),\label{eq:expansion1}
\end{eqnarray}
so that the normalization reduces to $\sum_{i}\left|c_{i}\right|^{2}=1$.
In the following we use an important property of the spherical harmonics $Y_{l}^{m}$, which reads $\cos(\nu)Y_{l}^{m}=\alpha_{l}^{m}Y_{l-1}^{m}+\alpha_{l+1}^{m}Y_{l+1}^{m}$ with the abbreviation $\alpha_{l}^{m}=\sqrt{\frac{(l-m)(l+m)}{(2l-1)(2l+1)}}$. With this we get
\begin{eqnarray}
\cos^{2}(\nu)Y_{l}^{m}=\alpha_{l}^{m}\alpha_{l-1}^{m}Y_{l-2}^{m}+\left[\left(\alpha_{l}^{m}\right)^{2}+\left(\alpha_{l+1}^{m}\right)^{2}\right]Y_{l}^{m}+\alpha_{l+2}^{m}\alpha_{l+1}^{m}Y_{l+2}^{m}.\label{eq:cos2}
\end{eqnarray}
Taking these properties in (\ref{eq:GP1}) into account leads to
\begin{eqnarray}
\varepsilon_{i}c_{i}+\lambda\sum_{j}M_{ij}c_{j}+4\pi g_{{\rm eff}}\sum_{jkl}\Gamma_{ijkl}c_{j}c_{k}c_{l}=\mu c_{i},\label{eq:GP2}
\end{eqnarray}
where we define the effective interaction strength $g_{{\rm eff}}=\bar{g}_{{\rm 2D}}N/4\pi A^{2}$. Furthermore, we introduce the abbreviations
\begin{eqnarray}
\varepsilon_{l} & = & \frac{\hbar^{2}l(l+1)}{2mA^{2}},\label{eq:def1}\\
M_{ij} & = & 2\pi\int_{0}^{\pi}d\nu\ \sin(\nu)Y_{i}^{0}(\nu)Y_{j}^{0}(\nu)\cos^{2}(\nu),\\
\Gamma_{ijkl} & = & 2\pi\int_{0}^{\pi}d\nu\ \sin(\nu)Y_{i}^{0}(\nu)Y_{j}^{0}(\nu)Y_{k}^{0}(\nu)Y_{l}^{0}(\nu),\label{eq:def2}
\end{eqnarray}
which have the specific values
\begin{eqnarray}
\Gamma_{ij00} =\frac{1}{4\pi}\delta_{ij}, \quad M_{i0}  =\frac{1}{3}\delta_{i0}+\frac{2}{3\sqrt{5}}\delta_{i,2,} \label{eq:def3}
\end{eqnarray}
where $\delta_{ij}$ denotes the Kroenecker delta.

Now we apply a generalization of Rayleigh-Schr\"{o}dinger perturbation theory \cite{Vrscay,Angyan1,Angyan2} in order to solve the nonlinear equation (\ref{eq:GP2}) perturbatively. To this end we employ the Taylor expansions of both the expansion coefficients $c_i$ and the chemical potential $\mu$ with respect to the dimensionless smallness parameter $\lambda$:
\begin{eqnarray}
c_{i} =c_{i}^{(0)}+\lambda c_{i}^{(1)}+\lambda^{2}c_{i}^{(2)}+\cdots, \quad \quad \mu =\mu^{(0)}+\lambda\mu^{(1)}+\lambda^{2}\mu^{(2)}+\cdots\, .
\end{eqnarray}
Considering the zeroth-order solution, we find $c_{i}^{(0)} =\delta_{i,0}$ and $\mu^{(0)} =g_{{\rm eff}}$.

In order to use the same normalization as in Rayleigh-Schr\"{o}dinger
perturbation theory, we follow Ref.~\cite{Angyan1} and
define the renormalization constant $Z(\lambda)$ according
to
\begin{eqnarray}
a_{i}=Zc_{i},\label{eq:c-a}
\end{eqnarray}
so we get
\begin{eqnarray}
\left|Z\right|^{2}=\sum_{i}\left|a_{i}\right|^{2}.\label{eq:renorm1}
\end{eqnarray}
The new coefficients $a_i$ are then normalized by $\sum_{i}a_{i}c_{i}^{(0)}=a_{0}=1$.
Furthemore, with this
Eq.~(\ref{eq:GP2}) becomes
\begin{eqnarray}
\varepsilon_{i}a_{i}+\lambda\sum_{j}M_{ij}a_{j}+\frac{4\pi g_{{\rm eff}}}{\left|Z\right|^{2}}\sum_{jkl}\Gamma_{ijkl}a_{j}a_{k}a_{l}=\mu a_{i},\label{eq:GP3}
\end{eqnarray}
 with the corresponding expansion 
\begin{eqnarray}
a_{i}=a_{i}^{(0)}+\lambda a_{i}^{(1)}+\lambda^{2}a_{i}^{(2)}+\cdots\, .\label{eq:expand2}
\end{eqnarray}
Note that here
the coefficients $a_{i}^{(l)}$ depend on $\left|Z\right|^{2}$. Therefore, we
treat $\left|Z\right|^{2}$ as another constant and expand the results
according to (\ref{eq:c-a}) and (\ref{eq:renorm1}) at the end. In
addition, since $a_{0}=1$, we have
\begin{eqnarray}
a_{0}^{(n)}=0,\qquad n>0,\label{eq:a01}
\end{eqnarray}
which implies that $\left|Z\right|^{2}=1+O(\lambda^{2})$. Thus, if one is interested only in first-order
perturbation theory, one could
consider
$\left|Z\right|^{2}$ equal to $1$.

Substituting (\ref{eq:expand2}) into (\ref{eq:GP3}) and applying
(\ref{eq:def3}), we get
\begin{eqnarray}
\varepsilon_{i}a_{i}^{(n)}+\sum_{j}M_{ij}a_{j}^{(n-1)}& + & 3\frac{g_{{\rm eff}}}{\left|Z\right|^{2}}a_{i}^{(n)}  +\frac{4\pi g_{{\rm eff}}}{\left|Z\right|^{2}}\sum_{\begin{array}{c}
{\tiny j,k,l}\\[-2mm]
{\tiny r,s,t<n}
\end{array}}\Gamma_{ijkl}a_{j}^{(r)}a_{k}^{(s)}a_{l}^{(t)}\delta_{r+s+t,n}\nonumber \\
& = & \mu^{(0)}a_{i}^{(n)}  +\sum_{m=1}^{n}\mu^{(m)}a_{i}^{(n-m)}.\label{eq:indices}
\end{eqnarray}
Considering $i=0$ in (\ref{eq:indices}), we obtain a recursion relation for the respective perturbative orders of the  chemical potential $\mu^{(n)}$
\begin{eqnarray}
\mu^{(n)}=\sum_{j}M_{0j}a_{j}^{(n-1)}+\frac{4\pi g_{{\rm eff}}}{\left|Z\right|^{2}}\sum_{\begin{array}{c}
{\tiny j,k,l}\\[-2mm]
{\tiny r,s,t<n}
\end{array}}\Gamma_{0jkl}a_{j}^{(r)}a_{k}^{(s)}a_{l}^{(t)}\delta_{r+s+t,n},\label{eq:req1}
\end{eqnarray}
while for $i>0$ in (\ref{eq:indices}), we get a recursion relation for the perturbative coefficients $a_{i}^{(n)}$
\begin{eqnarray}
a_{i}^{(n)} & = & \frac{1}{\varepsilon_{i}-\mu_{0}+\frac{3g_{{\rm eff}}}{\left|Z\right|^{2}}}\Bigg[-\sum_{j}M_{ij}a_{j}^{(n-1)}\nonumber\\
& & -\frac{g_{{\rm eff}}}{\left|Z\right|^{2}}\sum_{\begin{array}{c}
{\tiny j,k,l}\\[-2mm]
{\tiny r,s,t<n}
\end{array}}\Gamma_{ijkl}a_{j}^{(r)}a_{k}^{(s)}a_{l}^{(t)}\delta_{r+s+t,n}+\sum_{m=1}^{n}\mu^{(m)}a_{i}^{(n-m)}\Bigg]. \label{eq:req2}
\end{eqnarray}
Note that in the linear case, where we have $g_{{\rm eff}}=0$, Eqs.~\ref{eq:req1}) and (\ref{eq:req2}) reduce to the usual Rayleigh-Schr\"{o}dinger recursion formulae, where the normalization constant $Z$ does not appear. Furthermore, we read off from \ref{eq:req1}) and (\ref{eq:req2}) the first-order perturbative result
for $i>0$
\begin{eqnarray}
\mu^{(1)} & = & M_{00}=\frac{1}{3},\label{eq:mu1a}\\
a_{i}^{(1)} & = & -\frac{1}{\varepsilon_{i}-\mu_{0}+\frac{3g_{{\rm eff}}}{\left|Z\right|^{2}}}M_{i0}=-\frac{1}{\varepsilon_{2}+\frac{(3-\left|Z\right|^{2})g_{{\rm eff}}}{\left|Z\right|^{2}}}\frac{2}{3\sqrt{5}}\delta_{i2}. \label{eq:mu2b}
\end{eqnarray}
Since we have $\left|Z\right|^{2}=1+O(\lambda^{2})$, we conclude from (\ref{eq:c-a})
\begin{eqnarray}
c_{i} & =\delta_{i0}-\lambda\frac{1}{\varepsilon_{2}+2g_{{\rm eff}}}\frac{2}{3\sqrt{5}}\delta_{i2}+O(\lambda^{2}), \label{eq:coef1}
\end{eqnarray}
which gives for the ground state wave function according to (\ref{eq:expansion1})
\begin{eqnarray}
\xi_{0}(\nu) =\sqrt{\frac{N}{4\pi A^{2}}}\left[1-\lambda\frac{1}{\varepsilon_{2}+2g_{{\rm eff}}}\left(\cos^{2}\nu-\frac{1}{3}\right)\right]+O(\lambda^{2}).\label{eq:xi-final}
\end{eqnarray}
In addition we obtain from (\ref{eq:req1}) the second-order correction for the chemical potential:
\begin{eqnarray}
\mu^{(2)}=-\left(\frac{2}{3\sqrt{5}}\right)^2 \frac{1}{\varepsilon_{2}+2g_{{\rm eff}}}+3g_{{\rm eff}}\left(\frac{2}{3\sqrt{5}}\right)^2 \frac{1}{\left(\varepsilon_{2}+2g_{{\rm eff}}\right)^2} .\label{eq:mu2}
\end{eqnarray}

\begin{figure}
  \centering
  \includegraphics[width=\textwidth]{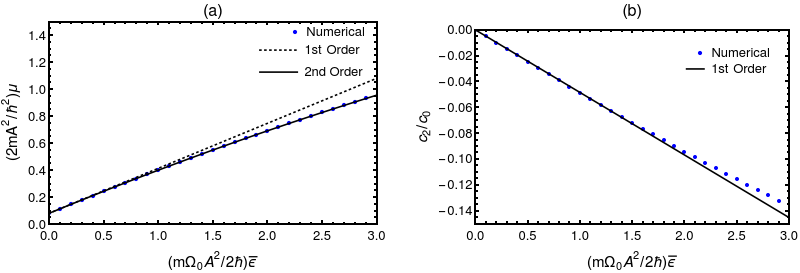}
  \caption{(a) Chemical Potential $\mu$ as a function of the surface eccentricity $\overline{\epsilon}$ for $2mN\bar{g}_{\rm 2D}/\hbar^2=1$. Black dashed and solid line represent analytical results corresponding to first- and second-order perturbation theory according to Eqs.~(\ref{eq:mu1a}) and (\ref{eq:mu2}), respectively. Blue dots are obtained by numerically by solving Eq.~(\ref{eq:GP2}). (b) Ratio between the coefficients $c_2$ and $c_0$ as a function of the surface eccentricity $\overline{\epsilon}$ according to Eq.~(\ref{eq:expansion1}) for $2mN\bar{g}_{\rm 2D}/\hbar^2=1$. The black solid line corresponds to the first-order perturbative result in Eq.~ (\ref{eq:coef1}), while blue dots are obtained from a numerical solution of Eq.~(\ref{eq:GP2}).}
\label{fig:stationary}
\end{figure}

In order to determine the range of validity for these perturbative results, we performed a comparative analysis, as depicted in Fig.~\ref{fig:stationary}. To this end we examined the chemical potential and the wave-function coefficients, as defined in Eq.~(\ref{eq:expansion1}), and contrasted them with the results obtained from a numerical solution of Eq. (\ref{eq:GP2}) using the relaxation method. Thus, we read off that first-order perturbative results are indistiguishable from the numerical ones for surface eccentricities 
$\overline{\epsilon} < 2 \hbar / m \Omega_0 A^2$.

\subsection{Perturbative excitation spectrum}
We consider now the dynamics of small excitations over the ground state $\xi^{(0)}$, which are described by the time-dependent GP equation 
\begin{eqnarray}
\frac{1}{2mA^{2}}\,\hat{L}^{2}\xi+\lambda\cos^{2}(\nu)\xi+\bar{g}_{{\rm 2D}}\left|\xi\right|^{2}\xi=i\hbar\,\frac{\partial\xi}{\partial t}\label{eq:TGP}
\end{eqnarray}
and the ansatz $\xi(t)=e^{-i\mu t/\hbar}[\xi_{0}+\delta\xi(t)]$. Here $\delta\xi(t)=ue^{-i\omega t}-\bar{u}^{\ast}e^{i\omega t}$
represent small elongations out of the ground state, yielding the Bogouliubov equations \cite{Pethick}
\begin{eqnarray}
TU=\hbar\omega\ \sigma U,\label{eq:BGE}
\end{eqnarray}
where we have introduced the abbreviations
\begin{eqnarray}
T & = & \left(\begin{array}{cc}
\frac{1}{2mA^{2}}\hat{L}^{2}+\lambda\cos^{2}(\nu)+2\bar{g}_{{\rm 2D}}\left|\xi_{0}\right|{}^{2}-\mu & -\bar{g}_{{\rm 2D}}\left|\xi_{0}\right|{}^{2}\\
-\bar{g}_{{\rm 2D}}\left|\xi_{0}\right|{}^{2} & \frac{1}{2mA^{2}}\hat{L}^{2}+\lambda\cos^{2}(\nu)+2\bar{g}_{{\rm 2D}}\left|\xi_{0}\right|{}^{2}-\mu
\end{array}\right),\label{eq:T}\\
\sigma & = & \left(\begin{array}{cc}
1 & 0\\
0 & -1
\end{array}\right),\\
U & = & \left(\begin{array}{c}
u\\
\bar{u}
\end{array}\right).
\end{eqnarray}
According to \cite{Peters, Carl}, we can apply a generalized version of Rayleigh-Schr\"{o}dinger perturbation theory to the generalized eigenvalue problem (\ref{eq:BGE}).

To this end, we must expand (\ref{eq:T}) in a power series of $\lambda$
\begin{eqnarray}
T=T_{0}+\lambda T_{1}+O\left(\lambda^{2}\right).
\end{eqnarray}
From (\ref{eq:mu1a}) and (\ref{eq:xi-final}), we have
\begin{eqnarray}
T_{0} & = & \left(\begin{array}{cc}
\frac{1}{2mA^{2}}\hat{L}^{2}+\frac{\bar{g}_{{\rm 2D}}N}{4\pi A^{2}} & -\frac{\bar{g}_{{\rm 2D}}N}{4\pi A^{2}}\\
-\frac{\bar{g}_{{\rm 2D}}N}{4\pi A^{2}} & \frac{1}{2mA^{2}}\hat{L}^{2}+\frac{\bar{g}_{{\rm 2D}}N}{4\pi A^{2}}\label{eq:TT0}
\end{array}\right),\\
T_{1} & = & \left(1-4\frac{\bar{g}_{{\rm 2D}}N}{4\pi A^{2}}\frac{1}{\frac{3\hbar^{2}}{mA^{2}}+\frac{\bar{g}_{{\rm 2D}}N}{2\pi A^{2}}}\right)\left(\cos^{2}\nu-\frac{1}{3}\right)\left(\begin{array}{cc}
1 & 0\\
0 & 1
\end{array}\right)\nonumber \\
 & + & 2\frac{\bar{g}_{{\rm 2D}}N}{4\pi A{{}^2}}\frac{1}{\frac{3\hbar^{2}}{mA^{2}}+\frac{\bar{g}_{{\rm 2D}}N}{2\pi A^{2}}}\left(\cos^{2}\nu-\frac{1}{3}\right)\left(\begin{array}{cc}
0 & 1\\
1 & 0
\end{array}\right).\label{eq:TT1}
\end{eqnarray}
Let us now express the problem in the basis of the spherical harmonics $\{Y_{l}^{m}\}$, i.e., 
\begin{eqnarray}
U(\nu,\phi)=\sum_{lm}c_{(l,m)}Y_{l}^{m}(\nu,\phi),
\end{eqnarray}
where the expansion coefficients are given by
\begin{eqnarray}
c_{(l,m)}= \int_{0}^{\pi}d\nu\sin\nu\int_{0}^{2\pi}d\phi\, Y_{l}^{m}(\nu,\phi)^{\ast} U(\nu,\phi).
\end{eqnarray}
Thus, the operator $T$ is represented in the basis of the spherical harmonics via the matrix elements
\begin{eqnarray}
T^{(l,m),(l^{\prime},m^{\prime})}= \int_{0}^{\pi}d\nu\sin\nu\int_{0}^{2\pi}d\phi\, Y_{l}^{m}(\nu,\phi)^{\ast}\, T \,Y_{l'}^{m'}(\nu,\phi).
\end{eqnarray}
%
From (\ref{eq:TT0}) we then deduce at zeroth order
\begin{eqnarray}
T_{0}^{(l,m),(l^{\prime},m^{\prime})}=\left(\begin{array}{cc}
\varepsilon_{l}+g_{{\rm eff}} & -g_{{\rm eff}}\\
-g_{{\rm eff}} & \varepsilon_{l}+g_{{\rm eff}}
\end{array}\right)\delta_{l,l^{\prime}}\delta_{m,m^{\prime},}\label{eq:T0}
\end{eqnarray}
while (\ref{eq:cos2}) and (\ref{eq:TT1}) yield the corresponding first-order contribution for $T$:
\begin{eqnarray}
T_{1}^{(l,m),(l^{\prime},m^{\prime})} & = & \left(1-4g_{{\rm eff}}\frac{1}{\varepsilon_{2}+2g_{{\rm eff}}}\right)\times \left(\begin{array}{cc}
1 & 0\\
0 & 1
\end{array}\right)\nonumber \\
 & \times & \left(\alpha_{l}^{m}\alpha_{l-1}^{m}\delta_{l,l^{\prime}-2}+\left[\left(\alpha_{l}^{m}\right)^{2}+\left(\alpha_{l+1}^{m}\right)^{2}\right]\delta_{l,l^{\prime}}+\alpha_{l+2}^{m}\alpha_{l+1}^{m}\delta_{l,l^{\prime}+2}-\frac{1}{3}\delta_{l,l^{\prime}}\right)\delta_{m,m^{\prime}}\nonumber \\
& + & 2g_{{\rm eff}}\frac{1}{\varepsilon_{2}+2g_{{\rm eff}}} \times \left(\begin{array}{cc}
0 & 1\\
1 & 0
\end{array}\right)\nonumber \\
& \times & \left(\alpha_{l}^{m}\alpha_{l-1}^{m}\delta_{l,l^{\prime}-2}+\left[\left(\alpha_{l}^{m}\right)^{2}+\left(\alpha_{l+1}^{m}\right)^{2}\right]\delta_{l,l^{\prime}}+\alpha_{l+2}^{m}\alpha_{l+1}^{m}\delta_{l,l^{\prime}+2}-\frac{1}{3}\delta_{l,l^{\prime}}\right)\delta_{m,m^{\prime}}.\label{eq:T1}
\end{eqnarray}
A direct diagonalization of (\ref{eq:T0}) yields for the eigenvalues for each $(l,\ m)$
\begin{eqnarray}
\hbar\left(\omega_{\pm}^{(l,m)}\right)_{0}=\pm\sqrt{\varepsilon_{l}(\varepsilon_{l}+2g_{{\rm eff}})},
\end{eqnarray}
with the corresponding eigenvectors
\begin{eqnarray}
\left(c_{(l^{\prime},m^{\prime})}^{(l,m)}\right)_{0}=\left(\begin{array}{c}
\frac{g_{\rm eff}+\varepsilon_{l}\pm\sqrt{\varepsilon_{l}(\varepsilon_{l}+2g_{{\rm eff}})}}{g}\\
1
\end{array}\right)\delta_{l,l^{\prime}}\delta_{m,m^{\prime}}.
\end{eqnarray}
\begin{figure}
  \centering
  \includegraphics[width=\textwidth]{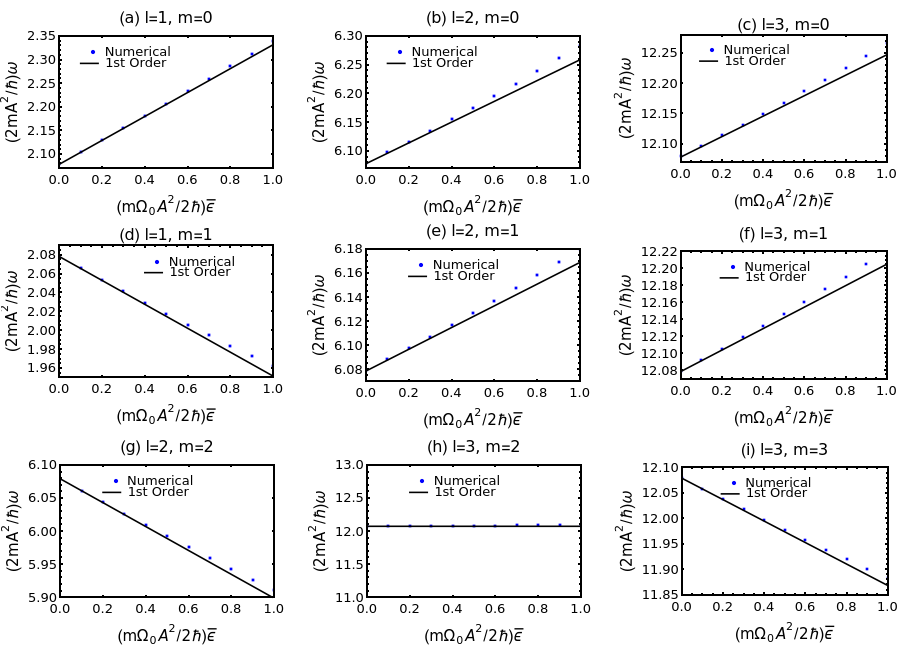}
\caption{Comparison of excitation frequencies between the first-order perturbative results in Eq. (\ref{eq:1st-order-omega}) and the eigenvalues obtained through direct numerical diagonalization of Eq. (\ref{eq:BGE}) for varying values of $l$ and $m$ for $2mN\bar{g}_{\rm 2D}/\hbar^2=1$. The figure demonstrates the validity range of the perturbative approximation and highlights an intriguing case for $l=3$ and $m=2$ (shown as Fig. \ref{fig:Plm}-h), where the first-order correction vanishes and the eccentricity of the trap has no effect on the corresponding eigenfrequency. }
  \label{fig:Plm}
\end{figure}
The first-order correction to the eigenvalues $\hbar\omega_{\pm}^{(l,m)}$ is then determined from the generalized Rayleigh-Schr\"{o}edinger theory, i.e.,
\begin{eqnarray}
\hbar\left(\omega_{\pm}^{(l,m)}\right)_{1}=\frac{\left<U_{\pm}^{(0)} \left|  T_{1} \right| U_{\pm}^{(0)}\right>}{\left<U_{\pm}^{(0)} \left| \sigma\right| U_{\pm}^{(0)}\right>},
\end{eqnarray}
which becomes in the basis of the spherical harmonics $\{Y_{l}^{m}\}$
\begin{eqnarray}
\hbar\left(\omega_{\pm}^{(l,m)}\right)_{1}=\frac{\sum_{(l',m'),(l'',m'')}\left(c_{(l'',m'')}^{(l,m)}\right)_{0}^{\dagger} T_{1}^{(l'',m''),(l',m')}\left(c_{(l^{\prime},m^{\prime})}^{(l,m)}\right)_{0}}{\sum_{(l',m')}\left(c_{(l',m')}^{(l,m)}\right)_{0}^{\dagger} \sigma \left(c_{(l^{\prime},m^{\prime})}^{(l,m)}\right)_{0}}.
\end{eqnarray}
Using (\ref{eq:T1}), we have concretely
\begin{eqnarray}
\hbar\left(\omega_{\pm}^{(l^{\prime},m^{\prime})}\right)_{1}=\pm \frac{\left[\left(\alpha_{l}^{m}\right)^{2}+\left(\alpha_{l+1}^{m}\right)^{2}-\frac{1}{3}\right]\left[ g_{\rm eff }\left( \varepsilon_2-2\varepsilon_l\right) + \varepsilon_l\varepsilon_2\right]}{\left(2g_{\rm eff}+\varepsilon_2\right)\sqrt{\varepsilon_l\left(\varepsilon_l+2g_{\rm eff}\right)}}.
\end{eqnarray}
With this, we obtain the final expression for the excitation frequencies:
\begin{eqnarray}
\omega_{\pm}^{(l,m)} & = & \pm \sqrt{\varepsilon_l\left(\varepsilon_l+2g_{\rm eff}\right)}
 \pm \lambda \frac{\left[\left(\alpha_{l}^{m}\right)^{2}+\left(\alpha_{l+1}^{m}\right)^{2}-\frac{1}{3}\right]\left[ g_{\rm eff }\left( \varepsilon_2-2\varepsilon_l\right) + \varepsilon_l\varepsilon_2\right]}{\left(2g_{\rm eff}+\varepsilon_2\right)\sqrt{\varepsilon_l\left(\varepsilon_l+2g_{\rm eff}\right)}}+O\left(\lambda^2\right). \label{eq:1st-order-omega}
\end{eqnarray}

In order to assess the validity range of the first-order perturbative result in Eq.~(\ref{eq:1st-order-omega}), we compare some eigenvalues as shown in Fig.~ \ref{fig:Plm} with the corresponding results obtained through direct numerical diagonalization of Eq.~(\ref{eq:BGE}). An intriguing behavior arises when considering the case of $l=3$ and $m=2$, which corresponds to Fig.~\ref{fig:Plm}-h. For these specific values of $l$ and $m$, the first-order correction in Eq. (\ref{eq:1st-order-omega}) vanishes, i.e.~changes in the eccentricity of the trap have no effect on the eigenfrequency. 
Thus, only higher-order corrections could potentially change the eigenfrequency. To date, we have no explanation for this occurrence, which is limited to such specific $l$ and $m$ values, nor have we determined if this result holds true for all eccentricity values. Further investigation is needed to comprehend this phenomenon. In addition, such results are consistent with  Refs. \cite{bubbleteo4,bubbleteo6} for $\bar{\epsilon}=0$, when their thin-shell limit is considered.

\section{Summary and conclusions}

In this paper, we have shown how the parameters of a bubble trap have to be carefully chosen in order to allow the existence of a thin-shell limit. Under the adequate assumptions, we were able to perform a dimensional compactification which leads to an effective 2D Hamiltonian for interacting BECs. From the experimental point of view, we demonstrated that, in order to avoid the collapse of the condensate, the thin-shell limit must be taken in such way that the spatial distortion caused by the eccentricity of the surface must be kept in the same order of magnitude as the width of the Gaussian distribution perpendicular to the surface. Details on how the experimentally obtained bubble-trap potentials fit into our general theory were also provided. In addition, applications of this theory to stationary systems as well as the excitation frequencies were calculated.
 
\section{Acknowledgments}

This work was supported by CNPq (Conselho Nacional de Desenvolvimento Cient{\' i}fico e Tecnol{\' o}gico), and DAAD-CAPES PROBRAL, Brazil Grant number 88887.627948/2021-00. N. S. M. acknowledges {\v S}tefan Schwarz Support Fund, projects OPTIQUTE APVV-18-0518, QUASIMODO VEGA 2/0156/22, and Grant No. 61466 from the John Templeton Foundation, as part of the ``The Quantum Information Structure of Spacetime (QISS)'' Project (qiss.fr). The opinions expressed in this publication are those of the author(s) and do not necessarily reflect the views of the John Templeton Foundation. N. S. M. thanks Nana Siddharth and Seyed Arash Ghoreishi for pointing out useful bibliography. Furthermore, A. P. acknowledges financial support by the Deutsche Forschungsgemeinschaft (DFG) via the Collaborative Research Center SFB/TR185 (Project No. 277625399).

\appendix

\section{Derivation of constants in $H_{0} $}\label{apendice}
It is necessary to specify some constants appearing for the Hamiltonian $H_{0} $. In order to do that, let us use the formula for the derivatives of the determinant of a quantity \textrm{B}
\begin{eqnarray}
\begin{cases}
\frac{d|B|}{d\lambda} =|B|\textrm{Tr}(B^{-1}\frac{dB}{d\lambda}),\\
\frac{d^{2}|B|}{d\lambda{{}^2}} =|B|\textrm{Tr}(B^{-1}\frac{dB}{d\lambda})^{2}+|B|\textrm{Tr}(B^{-1}\frac{d^{2}B}{d\lambda^{2}})-|B|\textrm{Tr}(B^{-1}\frac{dB}{d\lambda}B^{-1}\frac{dB}{d\lambda}).
\end{cases}
\end{eqnarray}
With this the metric $g$ has the following Taylor series up to second order
\begin{eqnarray}
g(s,\epsilon)=g_{0}+g_{1}\Lambda^{-1/2}y+g_{\epsilon}\Lambda^{-1}\bar{\epsilon}+\frac{1}{2}g_{2}\Lambda^{-1}y^{2}+O(\Lambda^{-3/2}),
\end{eqnarray}
where we have $g(0,0)=g_{0}$, $\frac{\partial g(0,0)}{\partial s}=g_{1}$, $\frac{\partial g(0,0)}{\partial \epsilon}=g_{\epsilon}$, and $\frac{\partial^{2} g(0,0)}{\partial s^{2}}=g_{2}$. Thus, the derivatives of $\gamma$ can be expressed as $\gamma_{1}  =\textrm{Tr}(g_{0}^{-1}g_{1})$ and $\gamma_{2} =\textrm{Tr}(g_{0}^{-1}g_{1})^{2}+\textrm{Tr}(g_{0}^{-1}g_{2})-\textrm{Tr}(g_{0}^{-1}g_{1}g_{0}^{-1}g_{1})$.
In the thin-shell limit the metric becomes the spherical one, yielding
\begin{eqnarray}
\hspace*{-0.5cm} 
g_{0}=\left(\begin{array}{ccc}
1 & 0 & 0\\
0 & A^{2} & 0\\
0 & 0 & A^{2}\sin^{2}\nu
\end{array}\right),\hspace*{3mm}
g_{1}=\left(\begin{array}{ccc}
0 & 0 & 0\\
0 & 2A & 0\\
0 & 0 & 2A\sin^{2}\nu
\end{array}\right),\hspace*{3mm}
g_{2}=\left(\begin{array}{ccc}
0 & 0 & 0\\
0 & 2 & 0\\
0 & 0 & 2\sin^{2}\nu
\end{array}\right).
\end{eqnarray}
This can be used to calculate the traces, granting $\gamma_{1}  =4/A$ and $\gamma_{2}=12/A^{2}$.\\

\end{document}